\documentclass[prd,preprint,eqsecnum,nofootinbib,amsmath,amssymb,
               tightenlines,dvips]{revtex4}
\usepackage{graphicx}
\usepackage{bm}

\begin {document}


\def\Mrowczynski{Mr\'owczy\'nski}
\def\Qs{Q_{\rm s}}

\def\alphas{\alpha_{\rm s}}
\def\hard{{\rm hard}}
\def\soft{{\rm soft}}

\def\D{{\bm D}}

\def\k{{\bm k}}

\def\ss{{\rm ss}}
\def\qmax{q_{\rm max}}

\def\p{{\bm p}}
\def\q{{\bm q}}

\def\v{{\bm v}}

\def\B{{\bm B}}
\def\A{{\bm A}}

\def\grad{{\bm\nabla}}

\def\etal{{\it et al.}}



\title
    {
     The turbulent spectrum created by non-Abelian plasma instabilities
    }

\author{Peter Arnold}
\affiliation
    {%
    Department of Physics,
    University of Virginia, Box 400714,
    Charlottesville, Virginia 22901, USA
    }%
\author{Guy D. Moore}
\affiliation
    {%
    Department of Physics,
    McGill University, 3600 University St.,
    Montr\'eal QC H3A 2T8, Canada
    }%

\date {September 21, 2005}

\begin {abstract}%
    {%
       Recent numerical work on the fate of plasma instabilities in
       weakly-coupled non-Abelian gauge theory has shown the
       development of a cascade
       of energy from long to short wavelengths.
       This cascade has a steady-state spectrum, analogous to the
       Kolmogorov spectrum for turbulence in hydrodynamics or for
       energy cascades in other systems.
       In this paper, we theoretically analyze
       processes responsible for this cascade and find a steady-state
       spectrum $f_\k \sim k^{-2}$,
       where $f_\k$ is the
       phase-space density of particles with momentum $\k$.
       The exponent $-2$ is
       consistent with results from numerical simulations.
       We also discuss implications of the emerging picture of
       instability development on the
       ``bottom-up'' thermalization scenario for
       (extremely high energy) heavy ion collisions,
       emphasizing fundamental questions that remain to be answered.
    }%
\end {abstract}

\maketitle
\thispagestyle {empty}


\section {Introduction}
\label{sec:intro}

It is important to understand theoretically the mechanisms by which
quark-gluon plasmas can locally equilibrate in heavy ion collisions such
as those at RHIC.  Since the path-breaking work of Baier, Mueller,
Schiff and Son
\cite{BMSS} on ``bottom-up'' thermalization,
one modest theoretical goal has been to understand the
process of equilibration in the simplifying theoretical limit of
arbitrarily high-energy collisions, where the running strong coupling
$\alpha_{\rm s}$ can be treated as small.  Typically, theorists expect
to be able to solve weakly-coupled problems, but equilibration of
weakly-coupled non-Abelian plasmas has proven to be a very rich and
challenging problem.  In particular, the original attempt by Baier 
\etal\ did not account for the physics of plasma instabilities,
which are now
believed to play a crucial role during
some of the early stages of thermalization \cite{ALM,shoshi,bodeker}.

Early in a heavy-ion collision, particles have an anisotropic
distribution of momenta, as measured in local frames moving with
the expanding plasma \cite{BMSS}. Later, they scatter and equilibrate
to locally isotropic distributions, giving rise to hydrodynamic
behavior.  Generically, anisotropic distributions of particles produce
collective plasma instabilities known as Weibel or filamentary
instabilities
\cite{weibel,heinz_conf,selikhov,pavlenko,mrow,mrow&thoma,strickland}.
These instabilities are associated with filamentation of particle
currents and initially exponential growth of long-wavelength magnetic fields.
Since these magnetic fields in turn deflect the particles in the plasma, and so
perhaps drive the particles towards isotropization and equilibration, it has
been important to understand just how large these unstable magnetic
fields grow \cite{AL,instability_prl,RRS,Dumitru,linear1,RRS2}.

In this paper, we will follow the nomenclature of Baier \etal\ and
refer to the initial particles of the plasma ({\it e.g.}\ small $x$
gluons in the saturation picture) as ``hard particles,'' and we will refer
to their momentum scale
({\it e.g.}\ the saturation scale) as $p_\hard$.
We will refer to the momentum scale of magnetic plasma
instabilities as $m_\soft$.  Once the system has expanded enough
that the hard partons have perturbative density
($n_\hard \ll p_\hard^3/g^2$), one generically finds that
$m_\soft \ll p_\hard$ \cite{ALM,instability_prl}.
We will assume and exploit this hierarchy of scales in what follows.
We will refer to any scale that is parametrically small
compared to $p_\hard$ as ``soft.''

\begin{figure}[t]
\includegraphics[scale=0.60]{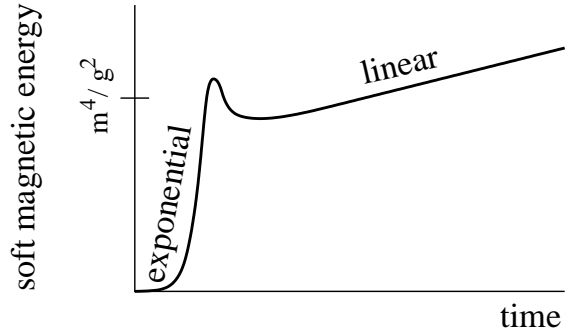}
\caption{%
    \label{fig:Benergy}
    A qualitative depiction of soft magnetic field energy vs.\/
    time for the growth of non-Abelian magnetic instabilities.
}
\end{figure}

Fig.\ \ref{fig:Benergy} shows a cartoon of the results of numerical
studies in 3+1 dimensions of non-Abelian plasma instability growth
\cite{linear1,RRS2} in a non-expanding system.  The energy in soft
magnetic fields is plotted versus time for simulations initiated with
a very tiny magnetic seed field.  The soft magnetic field energy
initially grows exponentially, but the behavior changes once the fields
become large enough that their non-Abelian self-interactions become
non-perturbative ($B^2 \sim m_\soft^4/g^2$).  The energy growth then
becomes linear with time.  For a non-expanding system, this linear
growth would continue until some parametrically long time later (not
addressed by the simulations) when the soft fields eventually have a
non-perturbatively large effect on the original, anisotropic hard
particles that created them.

\begin{figure}[t]
\includegraphics[scale=0.60]{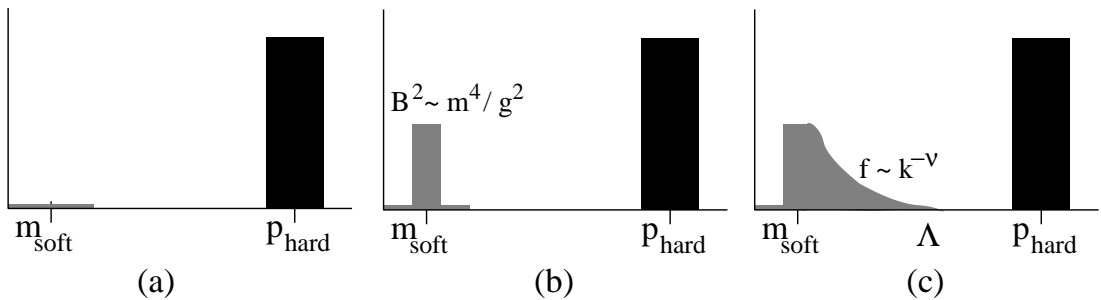}
\caption{%
    \label{fig:spec}
    A cartoon description of the energy spectrum of excitations at various
    times during the development of non-Abelian plasma instabilities.
}
\end{figure}

Recent numerical work \cite{linear2} indicates that the linear growth in
energy is associated with a cascade of energy from the soft,
perturbatively unstable modes toward the ultraviolet.  A cartoon of the
development of the spectrum is shown in Fig.\ \ref{fig:spec}.  (a) The
system starts with the initial, anisotropic hard particles of momentum
$\sim p_\hard$ plus some small initial fluctuations in softer modes.
(b) Plasma instabilities initially lead to rapid, exponential growth of
unstable soft modes of momenta $\sim m_\soft$.  (c) The unstable modes
stop growing in amplitude once they become non-perturbatively large, but
total soft energy continues to grow (linearly) as interactions
feed power from $m_\soft$
into increasingly higher momentum modes.  The spectrum of this cascade
falls with some characteristic power $k^{-\nu}$,
and the upper momentum limit $\Lambda(t)$ of the
cascade increases with time.  The goal of the present paper is to
theoretically determine the exponent $\nu$ by considering the processes
responsible for this cascade.

The basic plan of our analysis will be somewhat similar to
investigations of thermalization during reheating after inflation in the
early Universe \cite{son,tkachev1,tkachev}.  The details of our problem,
and the result for the spectral index $\nu$, are different, however.

Let $f_\k$ be the average occupancy of each mode $\k$.
In analyzing the cascade, we shall take
\begin {equation}
   1 \ll f_\k \ll \frac{1}{g^2}
\label {eq:fbounds}
\end {equation}
for $m_\soft \ll k \ll \Lambda(t)$.
In this range of $f$, the physics is perturbative ($f_\k \ll 1/g^2$) and
can be equally well described by a kinetic theory of particles or by
a classical theory of fields ($f_\k \gg 1$) \cite{MS}.%
\footnote{
  This is a slight oversimplification.  Perturbative occupation numbers
  $f_\k \ll 1/g^2$ are necessary but not sufficient for a kinetic theory
  description of relativistic systems.
  For quantitatively accurate (or qualitative
  order-of-magnitude) applications of
  kinetic theory,
  one must check
  that deBroglie wavelengths and the duration of scattering events
  are $\ll$ (or $\lesssim$)
  mean free paths, for whatever processes dominate the
  kinetic description.
}
The perturbative
nature of the cascade for $k \gg m_\soft$ follows from (i) the
non-perturbative occupation number $f \sim 1/g^2$ at the instability
scale $m_\soft$ and (ii) the decrease of $f$ as some power $k^{-\nu}$ to
be determined.  This gives
\begin {equation}
  f^\ss_\k \sim \frac{1}{g^2} \left( \frac{m_\soft}{k} \right)^\nu ,
\label {eq:fk}
\end {equation}
where the superscript ``$\ss$'' stands for ``steady-state.''
Perturbative cascades are known in the plasma
physics literature as ``weak plasma turbulence'' \cite{tsytovich}.
We will restrict attention
to the case where $\Lambda \ll g^{-2/\nu} m_\soft$
in order that (\ref{eq:fk}) satisfy the other condition $f_\k \gg 1$.
This classical field theory limit is implicit to the numerical
simulations which discovered the cascade \cite{linear1,RRS2,linear2},
since these simulations treated all soft fields as classical.

In the next section, we warm up by
discussing $2\to2$
scattering of particles at the UV end of the cascade.  We show that
this process alone would reproduce a result
for the spectral index $\nu$ in (\ref{eq:fk})
known from analogous scalar theory applications ($\nu=5/3$).
In Sec.\ \ref{sec:dominant}, we show that there is a more important
process: multiple scattering
from the non-perturbative infrared background produces a steeper
distribution, $\nu=2$.  This result is consistent with numerical simulations
\cite{linear2}.
Sec.\ \ref{sec:consistent} discusses the back reaction the
cascade has on the infrared physics of the instability and poses some
open questions concerning logarithmic effects in our analysis.
In Sec.\ \ref{sec:bottom-up}, we return to the question of how
instabilities affect the bottom-up scenario and discuss what remains to be
understood about instability development before this question can be answered.
Finally, we offer our conclusions.

Non-perturbative numerical studies of instability development in 3+1
dimensions have only studied hard particle distributions which are
moderately, but not extremely, anisotropic \cite{linear1,RRS2,linear2}.
In the case of moderate, $O(1)$ anisotropy, there is a single scale
which characterizes $m_{\rm soft}$, given parametrically by
\cite{ALM,mrow&thoma,strickland}
\begin {equation}
   m_\soft^2 \sim g^2 \int_\p \frac{f^\hard_\p}{p} \,,
\label {eq:msoft}
\end {equation}
where the integral is over the distribution of {\it hard}\/ particle
momenta $\p$.  For {\it extremely}\/ anisotropic distributions, the
situation is more complicated.  In most of this paper, we will
implicitly assume that there is only a single relevant soft scale.
We will only discuss the unresolved issues surrounding extremely anisotropic
distributions in our discussion of bottom-up thermalization in
Sec.\ \ref{sec:bottom-up}.


\section {Warm-up: Hard $2{\to}2$ scattering}
\label{sec:warmup}

The standard picture of driven turbulence is that something pumps energy
into the cascade from the infrared (IR) end at a constant rate.  In the
absence of dissipation, this leads to linear growth in the total energy.
At the same time, energy is flowing out the ultra-violet (UV) end by
increasing $\Lambda$.  The spectral index $\nu$ of the cascade can be
determined by balancing
(i) the rate at which energy flows out the UV end of the cascade by
scattering processes that extend the UV cascade by increasing $\Lambda$
({\it e.g.}\ scattering momentum $\Lambda$ particles to produce
momentum $2\Lambda$ particles) and
(ii) the constant rate energy enters the IR end of the cascade.

\begin{figure}[t]
\includegraphics[scale=0.40]{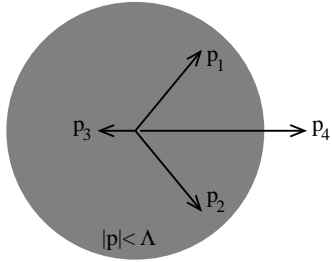}
\caption{%
    \label{fig:4momenta}
    Momenta in a $\p_1\,\p_2 \to \p_3\,\p_4$ process that creates a particle
    with momentum larger than $\Lambda$.
}
\end{figure}

As a review and warm-up example, let us for the moment (incorrectly) assume
that the dominant process for increasing the cut-off was
large-angle $2{\to}2$
scattering with all momenta roughly of order $\Lambda$.
Fig.\ \ref{fig:4momenta} shows an example of how two particles with
momenta $p_1$ and $p_2$ less than $\Lambda$ can produce a particle with
momentum greater than $\Lambda$ in $2{\to}2$ scattering.
Now consider what the time scale would be for bringing $f_{2\Lambda}$
up to its steady-state value (\ref{eq:fk}).  For a statistically
homogeneous system, the Boltzmann equation for $2{\to}2$ scattering
has the form
\begin {equation}
  \frac{d}{dt} \> f(p_4) =
  \int_{\p_1\p_2\p_3} |{\cal M}_{\p_1 \p_2 \leftrightarrow \p_3 \p_4}|^2
    \bigl[ f_1 f_2 (1+f_3)(1+f_4) - f_3 f_4 (1+f_1)(1+f_2) \bigr]
\end {equation}
for bosons such as the soft gluons that make up our cascade.
In the $f \gg 1$ classical field limit of (\ref{eq:fbounds}), this becomes
\begin {equation}
  \frac{d}{dt} \> f(p_4) =
  \int_{\p_1\p_2\p_3} |{\cal M}_{\p_1 \p_2 \leftrightarrow \p_3 \p_4}|^2
    \bigl[ f_1 f_2 (f_3+f_4) - f_3 f_4 (f_1+f_2) \bigr] .
\label {eq:boltz22}
\end {equation}
Even if $f(p_4)$ were initially small for $p_4 > \Lambda$ as in
Fig.\ \ref{fig:4momenta}, there is a term in this equation with
$f_1 f_2 f_3 \sim f_\Lambda^3$.  Take the square-amplitude
$|{\cal M}|^2$ for large-angle scattering
as order $g^4$ times normalizations and
the energy-momentum conserving delta function.
The characteristic rate $\Gamma$ implied by (\ref{eq:boltz22})
for $f(2\Lambda)$ to grow to its steady state value
$f^\ss_{2\Lambda} \sim f^\ss_\Lambda$ is then given by
\begin {equation}
  \Gamma f^\ss_\Lambda \sim g^4 (f^\ss_\Lambda)^3 \Lambda ,
\end {equation}
where the factor of $\Lambda$ on the right-hand side follows from
dimensional analysis.  So
\begin {equation}
  \Gamma \sim g^4 (f^\ss_\Lambda)^2 \Lambda .
\label {eq:Gamma22}
\end {equation}
From our assumed power-law form (\ref{eq:fk}) for $f^\ss$, this gives
\begin {equation}
  \Gamma \propto \Lambda^{1-2\nu} .
\label {eq:22exp1}
\end {equation}
Since the cutoff $\Lambda$ has to change on the time scale of the age
$t$ of the system, we must have $\Gamma \sim 1/t$.

On the other hand, let us estimate the rate from the linear growth of the
total energy density in the cascade.
Using (\ref{eq:fk}), the total energy density is
\begin {equation}
  \epsilon \sim \int d^3k \> k f_\k
  \sim \frac{\Lambda^4}{g^2} \left( \frac{m_\soft}{\Lambda} \right)^\nu
  \propto \Lambda^{4-\nu}
\end {equation}
(provided $\nu < 4$, which we will find).
Taking $\epsilon \propto t$, the corresponding rate is
\begin {equation}
  \frac{1}{t} \propto \Lambda^{\nu-4} .
\label {eq:IR}
\end {equation}
Comparing the exponent here to (\ref{eq:22exp1}), we would obtain
$\nu = 5/3$, in agreement with previous analysis in the literature
of certain cascades in field theory where large-angle $2{\to}2$
scattering is the dominant process \cite{tkachev}.%
\footnote{
  In particular, this $\nu$ agrees with Eq.\ (66) of Ref.\ \cite{tkachev}.
}

\section {The dominant process}
\label{sec:dominant}

The more efficiently the system can scatter energy into the UV, the
larger will be the exponent $\nu$.  The goal is to identify the most
efficient process.  We propose that the dominant process is multiple
scattering of quanta of momentum $\Lambda$ off of the non-perturbative
background field of wave-number $m_\soft$.  In this section, we analyze
this process and show that it is more important than the large-angle
$2{\to}2$ scattering considered in the last section.

\begin{figure}[t]
\includegraphics[scale=0.60]{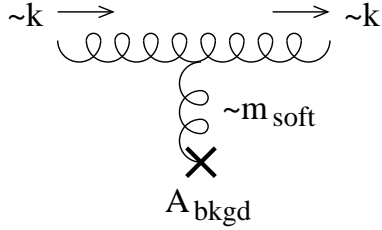}
\caption{%
    \label{fig:1to1}
    A particle with momentum $k \gg m_\soft$ scattering off of the
    non-perturbative field modes with wavenumber $m_\soft$.
}
\end{figure}

Consider a  relatively high-energy particle
($k \sim \Lambda \gg m_\soft$) propagating through a soft,
random, non-perturbatively large background field $A_{\rm bkgd}$
having wave-numbers of
order $m_\soft$.
The particle can scatter from the background as in Fig.\ \ref{fig:1to1}.
Non-perturbatively large means $g A_{\rm bkgd} \sim m_\soft$, and so
the only parameters in the problem are $k$ and $m_\soft$.
The probability of interaction is independent of $k$ in the
$k \gg m_\soft$ (Eikonal) limit.
By dimensional analysis,
the rate is therefore
$O(m_\soft)$ to change momentum by
$O(m_\soft)$.  Each such scattering will have a perturbative effect on
the particle, randomly changing its momentum $k$ by $m_\soft \ll k$.
$N$ successive collisions will change $k$ by order $N^{1/2} m_\soft$
in a time of order $N/m_\soft$. 
Taking $N \sim (k/m_\soft)^2 \sim (\Lambda/m_\soft)^2$, we see that
the time taken to change particle momenta from
$\Lambda$ to $2\Lambda$ at the UV end of the cascade would be
\begin {equation}
   t \sim \left( \frac{\Lambda}{m_\soft} \right)^2 \frac{1}{m_\soft} ,
\label {eq:t}
\end {equation}
corresponding to a net rate that depends on $\Lambda$ as
\begin {equation}
  \Gamma_{\Lambda{\to}2\Lambda} \sim \frac{1}{t} \propto \Lambda^{-2} .
\end {equation}
If we equate this to the rate (\ref{eq:IR}) determined by linear energy
growth, we obtain the spectral index
\begin {equation}
   \nu = 2 .
\end {equation}
The fact that this $\nu$ is larger than the $\nu=5/3$ of
Sec.\ \ref{sec:warmup} indicates that the large-angle $2{\to}2$
processes of Fig.\ \ref{fig:4momenta} are unimportant compared to
multiple scattering by Fig.\ \ref{fig:1to1}.%
\footnote{
  One can check this explicitly by comparing
  the rates for both $\nu=5/3$ and $\nu = 2$: Eq.\ (\ref{eq:Gamma22})
  gives $m_\soft (m_\soft/\Lambda)^{2\nu-1}$ for
  large-angle $2{\to}2$, and Eq. (\ref{eq:t}) gives in both cases a
  parametrically larger rate $m_\soft (m_\soft/\Lambda)^2$ for
  multiple scattering off of the non-perturbative background.
}

\begin{figure}[t]
\includegraphics[scale=0.60]{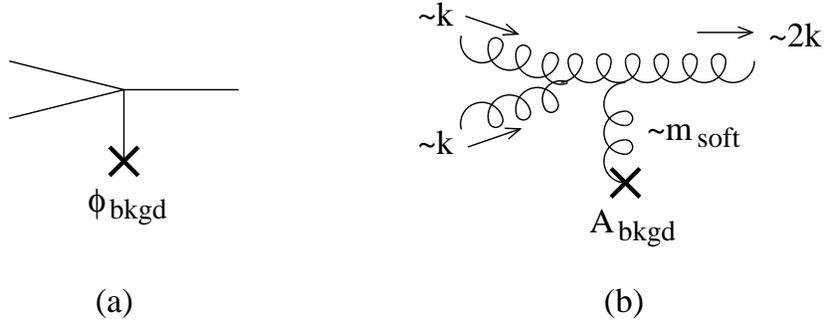}
\caption{%
    \label{fig:2to1}
    $2{\to}1$ processes catalyzed by a background field in
    (a) scalar $\phi^4$ theory and (b) gauge theory
    (inverse bremsstrahlung).
}
\end{figure}

We note in passing that interactions with a non-perturbative background
field have also been important in the study of cascades during reheating
of scalar field theories in the background of an oscillating inflation
field \cite{son,tkachev}.  However, processes analogous to Fig.\ \ref{fig:1to1}
were not relevant in that case because the background field carried zero
momentum and so could not directly transfer momentum to the particles.
In that case, the dominant process was
Fig.\ \ref{fig:2to1}a, which is not as efficient, giving
a smaller value of $\nu$ in those applications.
For our application, we consider in the Appendix the related process
of Fig.\ \ref{fig:2to1}b:
$2{\to}1$ inverse bremsstrahlung of particles
with momenta $k \gg m_\soft$, catalyzed by the
non-perturbative background field.
We find that, due to the Landau-Pomeranchuk-Migdal (LPM) effect, this
process is not competitive with multiple scattering via
Fig.\ \ref{fig:1to1}.

Failing to find a more efficient process than multiple scattering off
of the non-perturbative background, our provisional conclusion is
that the spectral index $\nu$ should be 2.  This is consistent with
results from simulations \cite{linear2}, which extract
$\nu = 2 \pm 0.2$
from Coulomb-gauge spectra of the cascade.  From (\ref{eq:IR}), we
also get that the UV end of the cascade should grow as
\begin {equation}
   \Lambda(t) \propto t^{1/(4-\nu)} \sim t^{1/2} .
\end {equation}
Using (\ref{eq:fk}), this means that our classical treatment
($f_k \gg 1$ for $k \ll \Lambda$) is valid for times
$t \lesssim 1/g^2 m_\soft$.


\section {Screening, scattering, and internal consistency}
\label {sec:consistent}

Having proposed a cascade with
\begin {equation}
   f^\ss(k) \sim \frac{1}{g^2} \left( \frac{m_\soft}{k} \right)^2 ,
\label {eq:fk2}
\end {equation}
we should now ask whether the development of the cascade might have
a back-reaction on the dynamics of the infra-red (IR) modes with
wave-numbers of order $m_\soft$.
These IR modes are the starting point for energy to enter the cascade,
and one could imagine that a back-reaction might conceivably change
the rate at which energy is taken into the cascade from the original
anisotropic hard particles.  If such a change to the rate depended on
$\Lambda(t)$ and so on time, it would be harder to understand why
the energy in the cascade grows linearly with time, as indicated by
numerical simulations.

Let us investigate how the cascade affects the soft scale $m_\soft$
itself, given by (\ref{eq:msoft}).  This scale
is initially created by the hard particle distribution, as
we have indicated explicitly in (\ref{eq:msoft}).
However, once the cascade develops, its
excitations could also contribute:
\begin {equation}
  \Delta m^2 \sim g^2 \int_\k \frac{f^{\rm cascade}_\k}{k}
  \sim g^2 \int_{\sim m_\soft}^{\sim \Lambda(t)} dk \> k \, f_k ,
\label {eq:dm}
\end {equation}
where $\k$ is integrated over momenta in the cascade.
How big is this contribution, and is its dependence on time through
$\Lambda(t)$ significant?
For $\nu<2$, the integral (\ref{eq:dm}) would be sensitive to
large $\Lambda(t)$; for $\nu>2$, it would not.
The spectral index $\nu=2$ found in this paper is the borderline case,
and (\ref{eq:fk2}) produces
\begin {equation}
  \Delta m^2 \sim m_\soft^2 \ln\left(\frac{\Lambda(t)}{m_\soft}\right) .
\label {eq:dm2}
\end {equation}

If we ignore the logarithm, then $\Delta m^2 \sim m_\soft^2$, and
the picture can be self-consistent.  The creation of the cascade would make
an $O(1)$ relative change to the IR physics driving the instability, and
so could make an $O(1)$ relative change to the rate of linear energy
growth.
That still allows the system
to settle down into steady growth as $t \to \infty$,
consistent with numerical results.

The formal logarithmic dependence of (\ref{eq:dm2}) on $\Lambda(t)$, however,
suggests that the energy growth rate should see logarithmic in time
corrections.
Current simulations \cite{linear2} do
not support such logarithmic corrections, perhaps because the
coefficient happens to be small, and perhaps because of the limited
statistical power of existing simulations.
In addition, the above analysis was rather crude: the details of
exactly how $\Delta m^2$ affects instability development depend,
for instance, on the degree of anisotropy of the cascade particles
as a function of time.%
\footnote{
  For instance, isotropic distributions do not produce magnetic
  instabilities.  Adding a large isotropic contribution $f^{\rm cascade}$
  to an anisotropic hard particle distribution $f^{\rm hard}$ will
  not change the momentum scale of the instability.  It will, however,
  decrease the perturbative growth rate of such instabilities.
  (See, for example, the discussion in Ref.\ \cite{bodeker}.)
  Roughly speaking, this is because
  changing magnetic fields require electric fields, the ability of
  a medium to support electric fields depends on its conductivity, and
  the conductivity is affected by the number of isotropic as well as
  anisotropic particles.
  On the other hand, if $f^{\rm cascade}$ is significantly anisotropic,
  it could increase the momentum scale of the instability, which would
  tend to increase growth rates.
  Numerical simulations suggest that the excitations in the cascade are
  somewhat but not perfectly isotropic.
  A simulation in Ref.\ \cite{linear2}, for
  a particular hard particle distribution, found that
  the magnetic fields making up the excitations in the cascade had
  $B_z^2$ equal to roughly 80\% of $B_x^2$ and $B_y^2$.
}
Thoroughly understanding the
physics of the cascade at the level of
logarithms remains an open problem.

We see that cascade particles can have an $O(1)$
effect on the infrared dynamics (up to logarithms).
In previous work \cite{boltzmann}, we argued that particles can make two
important contributions to the behavior of the plasma.
One is the effect on infrared fields (``screening,'' in the parlance
of Ref.\ \cite{boltzmann}).
The other is the contribution to the
scattering rate for hard particles.
That is, when a hard particle (or another cascade particle) scatters,
what fraction of the time is it scattering from a cascade particle?
This is determined by the relative contributions to the integral
\begin{equation}
\int_k f_\k (1 + f_\k) ,
\end{equation}
where $(1 + f_\k)$ is a final state Bose stimulation factor and where we
will focus on $k$'s in the cascade.
For our $f \propto k^{-2}$ cascade (with $f \gg 1$),
the integral is of form
\begin{equation}
\int^{\sim \Lambda}_{\sim m_\soft} k^{-4} k^2 dk \, ,
\end{equation}
which is dominated by the smallest momenta and insensitive to the moving
cutoff $\Lambda(t)$.  That is, the cascade particles are not important as
scatterers.  Most scattering is off the $m_\soft$ nonperturbative
background, not off the higher-momentum fluctuations.  This is another
way of seeing why it is scatterings from the soft background, rather
than between cascade particles, which determine the progress of the
cascade. 


\section {Whither Bottom-up thermalization?}
\label {sec:bottom-up}

We introduced this paper with a discussion of how nonabelian plasma
instabilities play a significant role in the bottom-up scenario
for quark-gluon plasma equilibration in the high energy limit.
We now have a reasonable picture of the development of non-Abelian
plasma instabilities in a non-expanding system.  Once we understand, at
least parametrically, the basic processes and
their rates, we should be able to apply this knowledge to
an analysis of expanding plasmas.  Unfortunately, there is an
obstacle.  For reasons of computational practicality, simulations
have been performed for moderately, but not {\it extremely},
anisotropic hard particle distributions.
For moderate anisotropy, there is a single parametric scale
$m_\soft$ for IR physics.
Non-perturbative IR magnetic fields are
therefore $O(m_\soft^2/g^2)$, the linear growth of soft
energy density is $O(m_\soft^5 t/g^2)$, and so forth.  However, as
we shall review, the first stages of the bottom-up scenario involve
{\it extremely}\/ anisotropic hard particle distributions, where
(in local fluid frames) the angle $\Delta\theta$ between typical
particle velocities and the transverse plane is parametrically small.
One must therefore understand the parametric dependence of
instability development on this additional small parameter, $\Delta\theta$.
We leave such understanding for future work.
In this section, we simply wish to emphasize the need
by showing how the current state of uncertainty can
have a significant impact on the first stage of bottom-up thermalization.
In particular, we will consider a possible alternative to the
recent analysis of the effects of instabilities by B\"odeker \cite{bodeker}.

In their original analysis of the stages of thermalization for
arbitrarily high energy heavy-ion collisions,
Baier \etal\ \cite{BMSS} considered an initial,
non-perturbatively dense state
of saturated low-$x$ gluons produced in the collision.
In their analysis, these were labeled the ``hard'' partons in the
plasma, with momenta $p_\hard$ of order the saturation scale $\Qs$.
Their analysis is restricted to following the process of thermalization
for $\tau \gg 1/\Qs$, where $\tau$ is proper time since the collision.

They assume initial conditions at time $\tau \sim 1/\Qs$ with
number density $n \sim \Qs^3/g^2$ of such hard particles.  This number
density dilutes as $n \sim (\Qs \tau)^{-1} \Qs^3/g^2$ due to the one
dimensional expansion of the system.  The scale (\ref{eq:msoft})
characterizing infrared physics (which they interpreted as a Debye
screening scale) then evolves with time as
\begin{equation}
m_\infty^2 \sim g^2 \int_\p \frac{f_\p}{p}
\sim \frac{g^2 n}{\Qs} \sim \Qs^2 (\Qs \tau)^{-1} \, .
\label{eq:m_Qs}
\end{equation}
Our notation ``$m_\infty$'' for this scale is in preparation for
our later discussion of plasma instabilities, for notational consistency with
previous work.%
\footnote{
  The historical reason for the subscript ``$\infty$'' is that,
  in addition to its role in the physics of instabilities,
  $m_\infty$ turns out to be
  the mass scale of transverse plasmons in the large momentum
  limit ({\it i.e.}\ $p \to \infty$) \cite{boltzmann}.
}

In the first phase of their analysis, the gluons dilute to
perturbative densities as the system expands, and their {\it local}
distributions of velocity temporarily become highly anisotropic due
to momentum selection: over time, approximately
free-streaming particles with the same $v_z$ would move to approximately
the same displacement $z$, along the beam axis,
from the origin of the collision.
This drive towards local anisotropy is moderated only by processes which
scatter the hard particles and so try to isotropize them.
Baier \etal\ considered random, individual, small-angle
$2{\to}2$ scattering of the hard particles and found that,
early in their scenario,
the balance
between expansion and scattering effects produces local distributions
with anisotropy
\begin {equation}
   \frac{p_z}{p_\perp} \sim \frac{1}{(\Qs \tau)^{1/3}}
\label {eq:pzBMSS}
\end {equation}
in local fluid rest frames.
This result indicates extreme anisotropy for
$\Qs \tau \gg 1$.
Later, however, once the system dilutes enough that
$f_\hard \lesssim 1$, Baier \etal\ found that other processes
come into play which eventually locally isotropize and thermalize the
system.  
Here, we modestly focus our attention on the
first phase, described by (\ref{eq:pzBMSS}) in the original scenario.%
\footnote{
   This corresponds to $1 \ll \Qs \tau \ll \alphas^{-3/2}$ in
   the original scenario of Ref.\ \cite{BMSS}.
}

Arnold, Lenaghan, and Moore \cite{ALM} pointed out that the first phase
will be modified by plasma instabilities.  Recently, B\"odeker
\cite{bodeker} has
attempted to deduce how scattering of hard particles from
soft fields created by plasma instabilities will
modify (\ref{eq:pzBMSS}).  B\"odeker assumed that soft fields
consist only of a non-perturbative component ($f_\soft \sim 1/g^2$)
with momenta $m_\soft \sim m_\infty$ given by (\ref{eq:msoft}).
For hard particle distributions with $O(1)$ anisotropy, this is
parametrically the only characteristic scale of unstable modes.
With these assumptions, B\"odeker found
\begin {equation}
   \frac{p_z}{p_\perp}
   \sim \frac{(m_\soft^3\tau)^{1/2}}{\Qs}
   \sim \frac{1}{(\Qs \tau)^{1/4}} \,,
\label {eq:pzB}
\end {equation}
which is parametrically larger than (\ref{eq:pzBMSS}).
This estimate is in keeping with the discussion of this paper; the
dominant way in which high momentum particles (whether in the cascade,
as we have discussed, or even higher momentum, as is relevant here)
undergo scattering, is by interaction with the nonperturbative fields at
the scale $m_\soft$.

However, this estimate is not secure, because the treatment of the
unstable scale $m_\soft$ has been made assuming $O(1)$ anisotropy, yet it
predicts a parametrically large anisotropy for times $\tau \gg 1/\Qs$.
Therefore we must reconsider the behavior of the instability for highly
anisotropic systems, a point also emphasized by B{\"o}deker
\cite{bodeker}.  As we shall review, there is another characteristic
scale $\qmax \gg m_\infty$ for such extremely
anisotropic distributions.
We believe that B\"odeker's estimate of taking $m_\soft$ of order
the lower scale ($m_\infty$) in (\ref{eq:pzB}) probably provides a
lower bound on the result for $p_z/p$.
We will not
attempt to make a complete treatment here, but will provide what we
believe is a sensible parametric upper bound.

For highly anisotropic, oblate distributions of hard particles,
the spread of angles of hard particle momenta $\p$ about the
transverse plane is characterized by
\begin{equation}
\Delta \theta \sim \frac{p_z}{p} \,.
\label{eq:deltatheta}
\end{equation}
The perturbatively
unstable modes are described by two scales.  One is $m_\infty$, defined
before, whose inverse
represents the time scale that a gauge field must act on
the particles before the back-reaction of particles on the field becomes
important.
$m_\infty$ characterizes the growth rate of instabilities.
The other scale is
the largest wave vector $q$
of any unstable soft gauge field mode, which is
\begin{equation}
\qmax \sim \frac{m_\infty}{\Delta \theta} \, .
\end{equation}

For a detailed explanation of how these scales arise, see Ref.\ \cite{ALM};
here we give a physical argument.  For a gauge field mode to be
unstable, the hard particles must fly for a time scale $\sim
m_\infty^{-1}$ in the presence of that mode, and be deflected in the
same direction that whole time.  That means that the magnetic field must
be roughly the same all along the trajectory of the particle's path.
Therefore, the gauge field must take roughly the same value over distances
$\delta x_\perp \sim 1/m_\infty$
in the two transverse directions, and over a distance
$\delta z \sim \Delta \theta/m_\infty$
in the longitudinal direction (since this
is how far a typical particle moves in the longitudinal direction in
that time scale).  That means that modes with $q_\perp \lesssim m_\infty$
and $q_z \lesssim m_\infty/\Delta \theta$ are unstable.  This agrees with
a more detailed analysis \cite{ALM}.

How large can the soft fields grow?  We can place an upper bound
as follows.  Gauge fields in different color
directions are all growing simultaneously.  When these grow large enough
that they can randomize the color of a hard excitation in a time scale
$\sim 1/m_\infty$, then the hard modes will no longer be contributing to
the time development of the unstable modes on the time scale required to
make the instability work.
The rotation of color charge as a hard particle moves along a trajectory is
a matrix given by the Wilson line
\begin {equation}
   {\cal P} \exp\left[ i g \int dx \cdot A \right] ,
\label {eq:rotate}
\end {equation}
where ${\cal P}$ represents path ordering.
In the analysis of a particular unstable mode, think of all the other
unstable modes as background.
The color rotation (\ref{eq:rotate}) due to a non-commuting
background field will therefore be
significant, and so stop instability growth,
if $g \, \delta x \cdot  A \gtrsim 1$.
To avoid color randomization of generic hard particles in our extremely
anisotropic distribution, we therefore need, for example,%
\footnote{
  In order to discuss parametric estimates concerning the gauge field
  $A$, one should
  assume that one is in a gauge where $A$ is relatively smooth.
  Here, we imagine that we study the state of the system at a
  particular time $t$ by (i) working in $A_0 = 0$ gauge, and then
  (ii) making a spatial gauge transformation to make $\A$ as smooth
  as possible at that particular time.
}${}^,$%
\footnote{
  Another way to arrive at the relation (\ref{eq:Aperp}) is to
  consider the propagator $(D_0 + \v\cdot\D)^{-1}$ of particle
  fluctuations in non-Abelian Vlasov equations.  In perturbative
  formulas for instabilities
  ({\it e.g.}\ Refs.\ \cite{ALM,mrow&thoma,strickland}),
  such propagators appear in the perturbative form
  $(\partial_0 + \v\cdot\grad)^{-1} \sim (\omega - \v\cdot\q)^{-1}$.
  One can then ask when the $g \v\cdot\A$ in $\v\cdot\D$ can be treated
  as a perturbation to $\omega$ and
  $\v\cdot\q \sim q_\perp + \Delta\theta \, q_z$, both of which are
  order $m_\infty$.
}
\begin {equation}
  A_\perp \lesssim \frac1{g \, \delta x_\perp} \sim \frac{m_\infty}{g} .
\label {eq:Aperp}
\end {equation}
Since the dominant instabilities have $\q$ nearly in the $z$ direction,
with $|\q| \sim \qmax$,
this implies that an {\it upper} bound on how large the corresponding
magnetic field strength can grow is%
\footnote{
  $q_\perp \sim m_\infty$.  Similar reasoning to (\ref{eq:Aperp})
  could allow $A_z \sim 1/(g \, \delta z) \sim \qmax/g$,
  which gives an $O(q_\perp A_z)$
  contribution to $B$ that is the same size as (\ref{eq:B}).
}
\begin {equation}
  B \sim \qmax A_\perp \sim
  \frac{\qmax m_\infty}{g} \sim \frac{m_\infty^2}{g \,\Delta\theta} \, ,
\label {eq:B}
\end{equation}
coherent on transverse length scales of $m_\infty^{-1}$.
Such fields will give momentum kicks to the hard modes due to the
$g \v\times \B$ term in the Lorentz force law.  The force is coherent
over the time scale $\delta t \sim m_\infty^{-1}$
(since this is both how often $B$ changes
as one moves in the transverse direction, and how often the particle's
charge gets rotated).  The individual momentum kicks are of size
$\Delta p_z \sim g B \, \delta t \sim \qmax$, and they lead
to momentum diffusion in the $p_z$ direction of size
\begin{equation}
p_z^2 \sim \frac{\tau}{\delta t} \, (\Delta p_z)^2
	\sim \qmax^2 m_\infty \tau .
\end{equation}
Therefore,
\begin {equation}
  \frac{p_z}{p_\perp} \sim \frac{(\qmax^2 m_\infty \tau)^{1/2}}{\Qs}
  \sim \frac{(m_\infty^3 \tau)^{1/2}}{\Qs \, \Delta\theta} \,.
\end {equation}
[Compare to (\ref{eq:pzB}).]
Substituting into the defining relation (\ref{eq:deltatheta})
for $\Delta \theta$,
and using the relation (\ref{eq:m_Qs})
for the time development of $m_\infty$, gives
\begin{equation}
\frac{p_z}{p_\perp}
\sim \sqrt{ \frac{(m_\infty^3 \tau)^{1/2}}{\Qs} }
\sim \frac1{(\Qs \tau)^{1/8}} \, .
\end{equation}

The above estimate is correct if our picture of how plasma instabilities
are cut off in a highly anisotropic setting proves correct.  While we
believe that our estimates represent an upper bound on such a cutoff, we
are not confident that they are correct.  
The moral of this story is that
better theoretical understanding is
needed of the non-perturbative dynamics of instabilities in the
case of extremely anisotropic distributions.
Until then, it will be difficult to analyze even the first phase
of bottom-up thermalization with complete certainty.


\section {Conclusions}

We have argued that, in the presence of non-Abelian instabilities that
have grown non-perturbatively large, the spectrum of the cascade of
soft field energy from unstable modes into the ultra-violet should
have the form $f_\k \sim k^{-\nu}$ with $\nu=2$.
This is consistent
with results from numerical simulations.

We should emphasize that this turbulent cascade is only a transitory
phenomenon appearing during the local thermalization of the quark-gluon
plasma in the high energy limit.  It is not directly related to the
spectrum of particles that much later leave the collision after
hadronization.
Also, though the word ``turbulence'' is used, the
existence of this cascade does not mean that the plasma is described by
ideal hydrodynamics at the time the cascade is formed.  At the early
times considered in this paper, most of the energy that will become the
quark-gluon plasma is still in hard particles, $p_\hard \sim \Qs$,
which have not yet thermalized and which are locally extremely
anisotropic.

It would be delightful to be able to apply the current understanding
of plasma instability development to produce a complete picture of
thermalization in the high energy limit, finally completing the
program initiated by the original bottom-up scenario.  A lot of
progress has been made over the last year in understanding the
qualitative behavior and parametric dependence of the non-perturbative
physics of instabilities for the case of moderately anisotropic
hard particles.  Unfortunately, figuring out the detailed role
of instabilities in bottom-up thermalization will likely require at least
a similar level of understanding of the case of {\it extremely}\/
anisotropic hard particles, which have not yet been investigated
non-perturbatively.


\begin{acknowledgments}

We would like to thank Larry Yaffe and Berndt Mueller for useful conversations.
This work was supported, in part, by the U.S. Department
of Energy under Grant Nos.~DE-FG02-97ER41027,
by the National Sciences and Engineering
Research Council of Canada, and by le Fonds Nature et Technologies du
Qu\'ebec.

\end{acknowledgments}


\appendix

\section{Inverse Bremsstrahlung}

In this appendix, we consider the $2{\to}1$ process of inverse
bremsstrahlung, catalyzed by the soft, non-perturbative field, as
depicted in Fig.\ \ref{fig:2to1}.  We will treat a case slightly more
general than that depicted in the figure, namely the case of one particle
with momentum $k\sim \Lambda$ and another particle with a momentum $k'$
larger than $m_\soft$, up to and including the scale $\Lambda$.
We will find that the coherence time
of this process is longer than the mean free time between
collisions of the particles with the soft background
(via Fig.\ \ref{fig:1to1}), and so we will have to account for the
Landau-Pomeranchuk Migdal (LPM) effect \cite{LPM} in estimating the
rate of inverse bremsstrahlung.

For an initial estimate, however, let us momentarily ignore the LPM
effect and consider the diagram of Fig.\ \ref{fig:2to1} in isolation.
Bremsstrahlung in a soft background is a small-angle, nearly-collinear
process.  Parametrically, the rate for bremsstrahlung or
inverse bremsstrahlung may be estimated as (i) the rate for small-angle
scattering from the background field (Fig.\ \ref{fig:1to1}),
times (ii) a factor of $g^2$ for absorbing
or emitting the additional gluon in Fig.\ \ref{fig:2to1}, times (iii)
an initial or final state factor of $f$ or $1+f$ for that gluon, and
(iv) a momentum integral $dk'/k'$ (responsible for the logarithmically
large rate of soft bremsstrahlung emission in vacuum).
As discussed in Sec.\ \ref{sec:dominant}, the rate for small-angle
scattering from the non-perturbative background is $O(m_\soft)$.
This gives
\begin {equation}
  \Gamma_{\mbox{\small(no LPM)}}
  \sim g^2 m_\soft \int \frac{dk'}{k'} \> f(k')
\label {eq:noLPM}
\end {equation}
for (inverse) bremsstrahlung.
To find the rate for a particle with momentum $k \sim \Lambda$
to substantially
change its energy, we must correct for two things.  First, absorbing a
momentum $k' \lesssim k$ makes only an order $k'/k$ relative change to $\k$.
Second, it is the difference between the rates
of absorption and emission of a particle of momentum $k'$ which is
relevant.  This involves the difference $f(k) - f(k+k')\sim f(k) \, k'/k$.
Therefore we must introduce two powers of $(k'/k)$ into the estimate
above, to get
\begin{equation}
\Gamma_{k\rightarrow 2k,\mbox{\small(no LPM)}}
\sim g^2 m_\soft \int \frac{dk'}{k'} \left(\frac{k'}{k}\right)^2 f(k')
\sim m_\soft \int_{m_\soft}^\Lambda \left( \frac{m_\soft}{k'} 
	\right)^\nu \: \frac{k' dk'}{k^2} \, ,
\label{eq:noLPM2}
\end{equation}
for doubling energy through inverse bremsstrahlung.
For $\nu \leq 2$, this is dominated by $k' \sim \Lambda$, and gives
$\Gamma \sim m_\soft^{1+\nu} \Lambda^{-\nu}$
(up to logarithms when $\nu=2$).
If we equated the power of $\Lambda$ in (\ref{eq:noLPM2}) with that in
(\ref{eq:IR}), we would obtain $\nu=2$, which would imply that this process
was parametrically just as important as the multiple soft scattering
treated in Sec.\ \ref{sec:dominant}.  However, the rate (\ref{eq:noLPM})
is an overestimate; we must account for the LPM effect.

In Abelian theories, LPM suppression occurs because (i) when a charged
particle scatters, the probability that a photon will be bremsstrahlung 
emitted is not very sensitive to the scattering angle, and (ii) an
emitted photon of a given wavelength
cannot resolve the difference between a single collision and $N$
collisions if the $N$ collisions are close enough together.%
\footnote{
  For a review of the LPM effect in photon bremsstrahlung, see
  Ref. \cite{spencer}. 
  For a discussion by the present authors of LPM suppression 
  in non-Abelian plasmas near equilibrium, see Refs.\ \cite{usLPM,boltzmann}.
  For a partial selection of earlier discussion of the LPM effect in
  non-Abelian gauge theories, see Refs.\ \cite{earlierLPM}.
}
The largest such $N$ can be determined parametrically by equating
(i) the time $N \tau$ for $N$ consecutive scatterings of the form of
Fig.\ \ref{fig:1to1}, where $\tau$ is the mean free time between
scatterings, with (ii) the formation time $1/(\delta E)$ for the
complete $1 \to 2$ bremsstrahlung (or $2 \to 1$ inverse bremsstrahlung)
process, where $\delta E$ is the off-shellness of the energy during
the process, which is
\begin {equation}
   \delta E = | E_1 + E_2 - E_* |
\label {eq:deltaE}
\end {equation}
for a $2{\leftrightarrow}1$ process with $\k_1\k_2 \leftrightarrow\k_*$.
For the reasons of resolution already mentioned, each $N$
collisions are equivalent to a single collision in terms of the
probability to induce bremsstrahlung, and so the rate $\Gamma$ for
(inverse) bremsstrahlung processes is a factor of $N$ smaller than
the naive estimate (\ref{eq:noLPM}) which treats every scattering
as independent.

The same sort of estimate works for gluon bremsstrahlung in non-Abelian
gauge theories.
Since bremsstrahlung is an almost collinear process, we can take
$k_\perp \ll k$, where $\perp$ is relative to the collinear axis,
and approximate (\ref{eq:deltaE}) for $k \gg m_\soft$ as
\begin {equation}
     \frac{m_1^2 + k_{1\perp}^2}{2 k_1}
   + \frac{m_2^2 + k_{2\perp}^2}{2 k_2}
   - \frac{m_*^2 + k_{*\perp}^2}{2 k_*} .
\label {eq:deltaE2}
\end {equation}
The main difference between photon bremsstrahlung in QED and
gluon bremsstrahlung in QCD is that, in QCD, any of
the three particles (including the bremsstrahlung gluon)
can undergo the scattering.
Therefore we may take
each $k_\perp^2 \sim N m_\soft^2$, in which case $\delta E$ is dominated
by the smallest energy particle, which is $k'$ in our case.  Further,
$m^2 \sim m_\soft^2 \ll k_\perp^2$ can be dropped, giving
\begin {equation}
   \delta E \sim \frac{k_\perp^2}{k'} \sim \frac{N m_\soft^2}{k'}.
\end {equation}
Plugging into (\ref{eq:deltaE2}), and equating the scattering time
$1/{\delta E}$ to $N \tau$ as outlined previously, we have
\begin {equation}
   N \tau \sim \frac{1}{\delta E} \sim \frac{k'}{N m_\soft^2} .
\end {equation}
Solving for $N$, and using the fact that $\tau \sim 1/m_{\soft}$ for
the individual scattering processes of Fig.\ \ref{fig:1to1}, determines
\begin {equation}
  N \sim \sqrt{k'/m_\soft} .
\end {equation}
Therefore, the rate is reduced by a factor of $(m_\soft/k')^{1/2}$,
which is a large suppression for $k' \gg m_\soft$.
This modifies Eq.\ (\ref{eq:noLPM2}) to,
\begin{equation}
\Gamma_{k\rightarrow 2k,\mbox{\small(LPM)}}
\sim m_\soft \int_{m_\soft}^\Lambda \left( \frac{m_\soft}{k'} 
	\right)^{(\nu+1/2)} \: \frac{k' dk'}{k^2} \, .
\end {equation}
This expression is dominated by small $k'$ if $\nu > 3/2$.  The $k'\sim
\Lambda$ edge of the range of integration provides
$\Gamma \sim \Lambda^{-\nu-1/2}$, which combined with 
(\ref{eq:IR}) would now
yield $\nu = 7/4$.  Therefore, {\em hard} bremsstrahlung is not a
competitive process.  However, the $k'\sim m_\soft$ edge of the
integration range gives $\Gamma \sim m_\soft^3/\Lambda^2$, which
combines with Eq.\ (\ref{eq:IR}) to give $\nu=2$.  Absorption of an
excitation with energy of order $m_\soft$ is competitive with (and
possibly indistinguishable from) scattering from such excitations, but
inverse bremsstrahlung of all harder scales is subdominant.



\end {document}